\documentstyle[emulateapj]{article}
\begin{document}

\title{Radio/X-ray Offsets of Large Scale Jets Caused by Synchrotron Time Lags}

\author{J.M. Bai\altaffilmark{1,2} and Myung Gyoon Lee\altaffilmark{1,3}}

\altaffiltext{1}{Astronomy Program, SEES, Seoul National University, Seoul 151-742, Korea;
jmmbai@astro.snu.ac.kr, mglee@astrog.snu.ac.kr}
\altaffiltext{2}{Yunnan Astronomical Observatory, NAO, Chinese Academy of Sciences, 
P.O. Box 110, Kunming, Yunnan 650011, China}
\altaffiltext{3}{Carnegie Institution of Washington, Department of Terrestrial Magnetism,
5241 Broad Branch Road, N.W., Washington, D.C., 20015}

\begin{abstract}

In the internal shock scenario, we argue that electrons in most kpc (or even larger) scale jets can be accelerated to
energies high enough to emit synchrotron X-rays, if shocks exist on these scales. 
These high energy electrons emit synchrotron radiation at high frequencies and cool as they propagate
downstream along the jet, emitting at progressively lower frequencies
and resulting in time lags and hence radio/X-ray 
(and optical/X-ray if the optical knot is detectable) offsets at bright knots,
with the centroids of X-ray knots being closer to the core.
Taking into account strong effects of jet expansion,
the behaviour of radio/X-ray and optical/X-ray offsets at bright knots
in M87, Cen A, 3C 66B, 3C 31, 3C 273, and PKS $1127-145$ is consistent with that of
synchrotron time lags due to radiative losses. This suggests that
the large scale X-ray and optical jets in these sources are due to synchrotron emission. 

\end{abstract}
\keywords{galaxies: active --- galaxies: jets --- radiation mechanism: nonthermal --- X-rays: galaxies}

\section{INTRODUCTION}

Since 1999, the {\it Chandra} X-ray Observatory has 
detected tens of X-ray jets in radio-loud AGNs. 
While the radio (and probably the optical) emission of the jets is 
certain to be synchrotron, the radiation mechanisms responsible for large scale 
X-ray jets have not been well understood yet (for a recent review, see Harris \& Krawczynski 2002). 
Although it seems clear that thermal models can be ruled out, 
both synchrotron and inverse-Compton emission models are possible.
For some sources, different models have been proposed by different research groups.
For example, the X-ray jet in 3C 273 can be interpreted
as synchrotron emission, or synchrotron self-Compton emission
or inverse-Compton scattering of the cosmic microwave background 
(CMB, Marshall et al. 2001; Sambruna et al. 2001; R\"oser et al. 2000).

Up to now, six large scale jets have been observed to have obvious
offsets between X-ray knots and the corresponding radio/optical knots, 
with the centroids of X-ray knots being closer to the core. 
They are M87 (Marshall et al. 2002; Wilson \& Young 2002), Cen A (Kraft et al. 2002), 
3C 273 (Marshall et al. 2001), PKS $1127-145$ (Siemiginowska et al. 2002), 
3C 66B (Hardcastle et al. 2001), and 3C 31 (Hardcastle et al. 2002).
In M87 (Sparks et al. 1996) and 3C 273 (Thomson et al. 1993), the centroids of 
optical knots are also closer to the core than their radio counterparts. 
The X-ray jets in M87, Cen A, 3C 31, and 3C 66B are interpreted as 
synchrotron emission of extremely high energy relativistic electrons, 
while the X-ray jet in PKS $1127-145$ is interpreted as 
inverse-Compton scattering of extremely low energy relativistic electrons with the CMB. 
But why in all these sources are knots at higher frequencies 
closer to the core than their counterparts at lower frequencies?
Could offsets between X-ray knots and radio/optical knots constrain the emission mechanisms of X-ray jets?

In blazars, similar phenomena have been observed among flares
which are generally believed to originate from the inner jets (pc and subpc scales, 
see e.g., Blandford \& K\"onigl 1979).
Multiwavelength monitoring of blazars shows that flares usually begin at high 
frequencies, then propagate to lower frequencies,
implying that high frequency synchrotron emission arises closer to the core than low frequency emission 
does (see e.g., Ulrich, Maraschi, \& Urry 1997; Marscher 1993, 2000).
For example, light curves of PKS $2155-304$ obtained in 1994 show that
an X-ray flare led a broader, lower amplitude extreme-ultraviolet flare by $\sim1$ day, 
and a broad, low-amplitude UV flare by $\sim2$ days (Urry et al. 1997).
Correlation analysis shows that 
a strong correlation appears between optical and radio (37 GHz) light curves,
with leading time of months or less (Tornikoski et al. 1994).
These leading flares are all due to synchrotron emission.
The similarity of offsets of knots on large scales to the time lags of flares 
on small scales thus may imply that the large scale X-ray jets in the above six sources 
are due to synchrotron emission.  

In an earlier paper (Bai \& Lee 2001), based on the knowledge of emission from the inner jets of blazars,
we predicted detectable synchrotron X-ray emission from most radio jets 
on kpc and a few 10 kpc scales (de-projected distance from the core). 
In this Letter, we argue that the radio/X-ray and optical/X-ray offsets of the above six jets are
caused by synchrotron time lags, 
and that X-ray emission in these large scale jets is synchrotron emission. 
We adopt $H_0=50$kms$^{-1}$Mpc$^{-1}$, and $q_{0}=0$ throughout this study.

\section{High Energy Electrons in inner jets and large scale jets}

In radio lobes, the energies of radio-emitting electrons are $\gamma\sim10^3$ to $\sim10^4$.
In inner jets, electrons of these energies should emit at optical/UV frequencies and
cannot travel to lobes during their radiative cooling time. 
They must be re-accelerated after leaving the inner jet. 

It is widely accepted that electrons are accelerated by internal shocks in the jet.
In the frame work of diffusive shock acceleration (e.g., Drury 1983; 
Blandford \& Eichler 1987), 
the time needed to accelerate electrons to energy $\gamma$ in the jet comoving frame is 
(Inoue \& Takahara 1996; Kusunose et al. 2000)
\begin{equation}
t_{acc}(\gamma) = \frac{20\lambda(\gamma)c}{3u_{s}^{2}} = \frac{20cm_e}{3e}\xi\gamma B^{-1},
\end{equation}
where $u_s\approx c$ is the shock speed, 
$B$ is the magnetic field strength, 
and $\lambda(\gamma) = \xi\gamma m_ec^2/(eB)$ is the mean free path of electrons
assumed to be proportional to the electron Larmor radius, with $\xi$ being a parameter
($\xi\geq1$, $\xi=1$ corresponds to the B\"ohm limit),
$m_e$ the electron mass, and $e$ the electron charge. 
The radiative cooling time $t_{cool}(\gamma)$ of relativistic electrons 
through synchrotron and inverse-Compton emission is
\begin{equation}
t_{cool}(\gamma) = \frac{3m_{e}c}{4\sigma_T}(U_B+U_r)^{-1}\gamma^{-1},
\end{equation}
where $\sigma_T$ is the Thomson cross-section, and $U_B = B^2/(8\pi)$ and $U_r$ are the energy densities 
of magnetic field and radiation (produced in or outside the jet), respectively.
Since $t_{acc}(\gamma)$ is proportional to $\gamma$ while
$t_{cool}(\gamma)$ is inversely proportional to $\gamma$,
at $t_{acc}(\gamma) = t_{cool}(\gamma)$, electrons are accelerated to
the maximum energy, $\gamma_{max}$ (Kirk et al. 1998; Kusunose et al. 2000).

Combining equations (1) and (2) yields
\begin{equation}
\frac{t_{acc}(\gamma)}{t_{cool}(\gamma)} = \frac{10\sigma_T(1+D)}{9\pi e}\xi B\gamma^2,
\end{equation}
where $D\equiv U_r/U_B$ is the ``Compton dominance" (e.g., Ghisellini et al. 1998).
In the jet comoving frame, the typical synchrotron emission frequency of relativistic electrons of $\gamma$,
averaged over pitch angles, is 
\begin{equation}
\nu_{s} = \frac{4}{3}\nu_{B}\gamma^{2} = \frac{2e}{3\pi m_{e}c}B\gamma^{2},
\end{equation}
where $\nu_B = eB/(2\pi m_{e}c)$ is Larmor frequency.
In terms of $\nu_s$, equation (3) can be expressed as
\begin{equation}
\frac{t_{acc}(\gamma)}{t_{cool}(\gamma)} = \frac{5\sigma_Tm_{e}c(1+D)}{3e^2}\xi\nu_{s}.
\end{equation}
Setting $t_{acc}(\gamma)/t_{cool}(\gamma)=1$ in equation (5), the maximum synchrotron frequency $\nu_{max}$
of $\gamma_{max}$ in the observer's frame is
\begin{equation}
\nu_{max} = \frac{3e^2(1+D)^{-1}}{5\sigma_Tm_{e}c}\xi^{-1}\frac{\delta}{1+z} \propto (1+D)^{-1}\xi^{-1}\delta,
\end{equation}
where $z$ is the redshift of the source, 
$\delta$ is the Doppler factor $\delta = [\Gamma(1-\beta\cos\theta)]^{-1}$, 
$\Gamma$ is the bulk Lorentz factor, and $\theta$ is the viewing angle of the jet.

The kpc scale jet is an extension of the inner jet, and is still relativistic, as indicated by 
the observed jet$-$counterjet intensity asymmetry. Therefore,
$\xi$ and $\delta$ do not change much along the jet from
small scales to kpc scales, unlike 
$B$ which decreases quickly along the jet. 
On large scales, from subkpc to tens of kpc, the dominant cooling process in the jets 
is synchrotron emission (Celotti et al. 2001), i.e., $1+D_{kpc} \sim 1$. 
Thus, according to equation (6),
$\nu_{max}({\rm kpc}) \sim (1+D_{pc})\nu_{max}({\rm pc})$.
In the inner jets of some sources, such as Mrk 501, X-rays are due to synchrotron emission, and  
the dominant cooling process is synchrotron cooling, i.e., $1+D_{pc} \sim 1$, so
$\nu_{max}$ is roughly constant along the jet (though 
$\gamma_{max}$ is much higher on kpc scales), and 
electrons can be re-accelerated to energies high enough to emit synchrotron X-rays on kpc or even larger scales
if shocks exist on these scales.

In the inner jets of OVV quasars and probably FR II radio galaxies,
the synchrotron peak frequencies are in the IR/optical bands, and 
the dominant cooling process is inverse-Compton emission. In these sources,
$D_{pc}$ is typically in the range of 10 -- 100, 
and as much as 1000 in some luminous quasars (Urry 1999). It is obvious that
in some sources, the maximum synchrotron frequencies and hence the peak frequencies
in the kpc scale jets are more than 100 times larger than those in
the inner jets, and that X-rays are dominated by synchrotron emission.
That is to say,
electrons in kpc scale jets of these sources 
can be accelerated to energies high enough to produce synchrotron X-ray jets if shocks exist on these scales.

Therefore, shocks in most large scale jets can re-accelerate electrons to energies  
high enough to emit synchrotron X-rays.
It is not strange that electrons in large scale jets can be accelerated 
to energies higher than the maximum energy in inner jets. 
Even the optical-emitting electrons in some large scale optical knots
have higher energies than the maximum energy in the inner jets.
For example, in 3C 273, 
the energies of optical electrons in knot D and H are $\gamma>10^5$ (R\"oser \& Meisenheimer 1991), 
much larger than the maximum energy of $\gamma\sim 10^4$ in the inner jet (Ghisellini et al. 1998).

\section{Offsets of knots in large scale jets}

It can be seen in equation (4) that
high energy electrons emit synchrotron radiation at high frequencies and cool, 
emitting at progressively lower frequencies and resulting in time lags
between high ($\nu_1$) and low ($\nu_2$) frequencies. 
The time lag of emission at $\nu_2$ ($\gamma_2$) to emission at $\nu_1$ ($\gamma_1$) is
the time for electrons to lose energy $\Delta\gamma =\gamma_1 - \gamma_2$, i.e.,
\begin{equation}
t_{lag} = \int_{\gamma_1}^{\gamma_2}\frac{d\gamma}{\dot{\gamma}}= \frac{-3m_{e}c}{4\sigma_T(U_B+U_r)}\int_{\gamma_1}^{\gamma_2}\frac{d\gamma}{\gamma^2},
\end{equation}
where $\dot{\gamma} = -4\sigma_T(U_B+U_r)\gamma^2/(3m_{e}c)$ is the cooling rate at $\gamma$,
and $t_{lag}$ is in the jet comoving frame.
Integration yields
\begin{equation}
t_{lag} =\frac{3m_{e}c(\gamma_2^{-1}-\gamma_1^{-1})}{4\sigma_T(U_B+U_r)} = t_{cool}(\gamma_2) -t_{cool}(\gamma_1).
\end{equation}
That is to say, the time lag of synchrotron emission at $\nu_2$ to that at $\nu_1$ 
is equal to the difference between the cooling time of electrons of energies of $\gamma_2$ and $\gamma_1$. 
In terms of $\nu_1$ and $\nu_2$, $t_{lag}$ is
\begin{equation} 
t_{lag}  = \frac{2\sqrt{6\pi em_{e}c}}{\sigma_T(1+D)}B^{-\frac{3}{2}}(\nu_2^{-\frac{1}{2}}-\nu_1^{-\frac{1}{2}}).
\end{equation} 
This is also the time lag between the peaks (not the beginnings) of synchrotron flares at $\nu_2$ and $\nu_1$, 
provided that the light-crossing time is unimportant relative to the electron injecting time, 
and that $\gamma \ll \gamma_{max}$ 
(e.g., Zhang et al. 2002; Chiappetti et al. 1999; Chiaberge \& Ghisellini 1999;
Georganopoulos \& Marscher 1998). 
If $\nu_1\gg \nu_2$, 
then $t_{cool}(\nu_2)\gg t_{cool}(\nu_1)$, and
$t_{lag} \approx t_{cool}(\nu_2)$,  
i.e., the time lag of a flare at low frequency to its corresponding synchrotron 
flare at high frequency 
is roughly the radiative cooling time of relativistic electrons at lower frequency (Urry et al. 1997, 1999).

In the case that X-rays are due to synchrotron emission, 
synchrotron cooling is the dominant cooling process of electrons, i.e., $1+D\sim 1$.
Thus, in the jet comoving frame
\begin{equation} 
t_{lag}\approx t_{cool}(\nu_2) = 2\times 10^4[(1+z)/\delta]^{-\frac{1}{2}}B^{-\frac{3}{2}}\nu_{15}^{-\frac{1}{2}}  \hspace*{1mm} {\rm s},
\end{equation}
where $B$ is in gauss, $\nu_{15}\equiv \nu_{2}/(10^{15} {\rm Hz})$ is in the observer's frame, 
and $t_{lag}$ is in units of second. 
The time lags of optical (V band, observer's frame) emission to 
synchrotron X-ray emission in the jet comoving frame are 
\begin{equation}
t_{o/x} \approx 1.477\times 10^4[(1+z)/\delta]^{-\frac{1}{2}}B^{-\frac{3}{2}}  \hspace*{1mm} {\rm s},
\end{equation}
and the time lags of radio (1 GHz, observer's frame) emission to X-ray emission in the jet comoving frame are 
\begin{equation}
t_{r/x} \approx t_{r/o} \approx 2\times 10^7[(1+z)/\delta]^{-\frac{1}{2}}B^{-\frac{3}{2}}  \hspace*{1mm} {\rm s}.
\end{equation}
As the emitting plasma propagates downstream along the jet, $t_{o/x}$ and $t_{r/x}$
result in offsets between the centroids of optical and X-ray knots, 
and between the centroids of radio and X-ray knots, respectively,
with the X-ray knot being earlier and closer to the core than radio and optical knots, 
and optical knot than radio knot. 
The radio/X-ray or optical/X-ray knot pairs observed at the same time 
are thus actually produced by two different populations of electrons, as
pointed out by Siemiginowska et al. (2002). 
However, bright knots on large scales are sites continuously 
generating shocks and re-accelerating electrons, as indicated by proper motion at knots 
of M87 jet (Biretta et al. 1995, 1999).
The magnetic field strength in bright knots and the jet speed $v_j$
are roughly constant (on time scale of years). 
Offsets produced by a series of populations of electrons at a bright knot 
are approximately the same. 
That is to say, the observed radio/X-ray and optical/X-ray offsets 
at a bright knot still reflect the synchrotron time lags, 
as if they were produced by a single population of electrons at the same time.

Assuming $v_j=0.42c$ ($\Gamma=1.1$, Celotti et al. 2001) on large scales,
the observed offset between optical (V band) and X-ray knots caused by $t_{o/x}$ simply is 
\begin{equation}
D_{o/x} \approx 59.77B_{-4}^{-\frac{3}{2}}\Gamma[(1+z)/\delta]^{-\frac{1}{2}}\sin\theta \hspace*{1mm} {\rm pc},
\end{equation}
where $B_{-4}\equiv B/(10^{-4} {\rm gauss})$. 
Similarly, the observed offsets between radio (1 GHz) and X-ray knots caused by $t_{r/x}$ is
\begin{equation}
D_{r/x} \approx D_{r/o} \approx 80.95B_{-4}^{-\frac{3}{2}}\Gamma[(1+z)/\delta]^{-\frac{1}{2}}\sin\theta \hspace*{1mm} {\rm  kpc}.
\end{equation}

The typical magnetic strength in large scale jets is $B\sim 10^{-4}$ gauss (Harris \& Karwczynski 2002).
According to equations (13) and (14), the de-projected optical/X-ray and radio/X-ray offsets of large scale jets
are $\sim 65.7$ pc and $\sim 89$kpc, respectively.
However, jet expansion on large scales 
may shorten the synchrotron time lags and hence the offsets, especially radio/optical offsets. 
Lower energy electrons cool slower, hence at lower frequencies, 
jets have longer time to expand sideways and thus are wider.
Multiband imaging of jets in Pictor A (Wilson et al. 2001), Cen A (Kraft et al. 2002),
M87 (Marshall et al. 2002; Sparks et al. 1996), 
and 3C 273 (Thomson et al. 1993) shows that radio jets are much wider, 
indicating that jet expansion on large scales is significant.
It is reasonable that the magnetic field farther away from the jet axis is weaker. 
When the jet expands, some electrons move sideways to region with weaker magnetic field, 
emitting at lower frequencies earlier than they should. 
In other words, jet expansion reduces the time lags for these electrons, 
as well as the jet emission at some frequencies. 
On large scales, the magnetic field is very weak, and
only at frequencies close to $\nu_{max}$, is the jet expansion negligible.
At low frequencies (in some cases may be as high as optical), the jet may expand so greatly that 
the emission of the jet is mainly contributed by electrons in the expanded region of the jet,
and consequently the centroids of emission at these frequencies shift upstream, i.e.,
the offsets to high frequency knots are shortened. 

In equation (9), it can be seen that at a given frequency, synchrotron time lags and thus offsets 
between two emission frequencies are inversely proportional to the magnetic field strength $B$.
Because $B$ decreases along the jet,
offsets due to synchrotron time lags should increase along the jet provided that 
jet expansion is unimportant. 
In M87, Cen A, 3C 273 and PKS $1127-145$ more than two knots have been detected. 
The offsets in M87, Cen A and 3C 273 indeed get larger along the jet, 
suggesting that these offsets are due to synchrotron time lags. 
In M87, from knot A to Knot B, the optical/X-ray offsets increase 
from $0.09 \pm 0.02\arcsec$ to $0.29 \pm 0.06\arcsec$, 
and from the core to knot A, 
the optical/X-ray offsets are too small to be measured at current accuracy of 0.03$\arcsec$ 
(Marshall et al. 2002). 
The radio/optical offsets in M87 also get larger along the jet, 
from $\sim0.24\pm 0.01\arcsec$ at knot HST-1 to $\sim0.3\pm 0.01\arcsec$ at knot D.
In Cen A, the radio/X-ray offsets increase 
from $<0.5\arcsec$ at A1/AX1 knots and $\sim2.5\arcsec$ at A2/AX2 knots
to $\sim5\arcsec$ at knot B (Kraft et al. 2002). 
In 3C 273, the optical/X-ray offsets increase 
from $\sim0.22\arcsec$ at knot A to $\sim0.5\arcsec$ 
at knot B (see Fig. 3 in Marshall et al. 2001).
The radio/X-ray offsets in PKS $1127-145$ and the radio/optical offsets 
beyond knot D in M87 decrease along the jet, 
but this may be due to strong expansion of the jet.

In fact, from pc and subpc scales to large scales, offsets also increase along the jet.
The typical time lag of optical flares to X-rays flares is of day scale, 
such as in PKS $2155-304$ (Urry et al. 1997) and Mrk 421 (Buckley et al. 1996).
The corresponding optical/X-ray offsets on pc and subpc scales 
are of the order of lightday, i.e. thousandths of pc. 
On large scales, 
the observed optical/X-ray offsets in large scale jets are of the order of tens (M87, assuming $\theta\sim 30\arcdeg$) 
to hundreds (3C 273 and 3C 66B) of pc. 
The typical time lag of radio flares to optical flares is of the scale of months (e.g., Tornikoski et al. 1994), 
corresponding to radio/optical and radio/X-ray offsets of the order of 0.1 pc on pc and subpc scales, 
while the observed radio/X-ray offsets on large scales are of the order of tens and hundreds of pc (Cen A, M87 and 3C 31) to tens
of kpc (PKS $1127-145$). 
These also suggest that the observed radio/X-ray and optical/X-ray offsets in the six large scale jets are due to
synchrotron time lags. 

According to equations (13) and (14), the optical/X-ray offset at a knot should be 
$\sim 10^{-3}$ times smaller than the radio/optical offset at the same knot. 
In M87, though shortened due to jet expansion, 
the radio/optical offsets of $\sim0.24\pm 0.01\arcsec$ at knot HST-1 and $\sim0.3\pm 0.01\arcsec$ at knot D
are larger than the corresponding optical/X-ray offsets 
which are too small to be measured at current accuracy of 0.03$\arcsec$ 
(Marshall et al. 2002). 
This confirms that offsets in M87 are due to synchrotron time lags. 
Among the six sources, M87 is the only one that has data for both optical/X-ray and radio/optical offsets.
Probably, the optical/X-ray offsets in the rest five sources were also
smaller than the corresponding radio/optical offsets, if they could be measured.

\section{Discussion and Conclusions}

It should be pointed out that, unlike bright knots, a weak knot may be caused by a single shock.
Weak X-ray knots thus may not have corresponding optical or radio knots before disappearing.
The faint knots, A4/AX4, A5/AX5 and A6/AX6 in Cen A, and Dx and F in M87 are probably a single shock wave,
and the offsets at these knots may not be caused by synchrotron time lags. 

In some sources, the large scale X-ray jets may be 
dominated by inverse-Compton scattering of radio-emitting or even lower energy 
electrons (Schwartz et al. 2000, Schwartz 2002; Tavecchio et al. 2000; Celotti et al. 2001).
In these sources, 
X-ray knots cannot be closer to the core than the corresponding radio knots, 
as indicated by multiband monitoring of flares in PKS $1510-10$ in which
X-rays are due to inverse-Compton scattering 
and the X-ray flare in 1997 lagged behind the radio flare $\sim 16$ days (Marscher 2000).
In the six sources, because the bright X-ray knots 
precede their radio (and optical) counterparts,
the X-ray jets cannot be due to inverse-Compton emission.

In conclusion, 
shocks in most large scale jets can re-accelerate electrons to energies  
high enough to emit synchrotron X-rays;
The observed radio/X-ray and optical/X-ray offsets at bright knots 
in M87, Cen A, 3C 66B, 3C 31, 3C 273, and PKS $1127-145$ are probably caused by synchrotron time lags,
suggesting that the large scale X-ray and optical jets in these sources are dominated by synchrotron emission.

We thank the anonymous referee and D.E. Harris for constructive suggestions, comments and discussions.
This work was financially supported by the BK21 Project of the Korean government.

\end{document}